\documentclass[aps,prl,twocolumn,superscriptaddress,showpacs]{revtex4}
\usepackage{amssymb}
\usepackage{txfonts}
\usepackage{tipa}
\usepackage{graphicx}% Include figure files
\usepackage{dcolumn}% Align table columns on decimal point
\usepackage{bm}% bold math

\begin{document}

%\preprint{APS/123-QED}

\title{Magnetic order in CaFe$_{1-x}$Co$_x$AsF (\emph{x} = 0, 0.06, 0.12) superconductor compounds}% Force line breaks with \\

\author{Y. Xiao}
\email[Electronic address: y.xiao@fz-juelich.de]{}
%\thanks{}
\affiliation{Institut fuer Festkoerperforschung, Forschungszentrum
Juelich, D-52425 Juelich, Germany}

\author{Y. Su}
%\email[]{}
%\thanks{}
\affiliation{Juelich Centre for Neutron Science, IFF,
Forschunszentrum Juelich, Outstation at FRM II, Lichtenbergstr.
1,D-85747 Garching, Germany}

\author{R. Mittal}
%\email[]{}
%\thanks{}
\affiliation{Juelich Centre for Neutron Science, IFF,
Forschunszentrum Juelich, Outstation at FRM II, Lichtenbergstr.
1,D-85747 Garching, Germany}

\affiliation{Solid State Physics Division, Bhabha Atomic Research
Centre, Trombay, Mumbai 400 085, India}

\author{T. Chatterji}
%\email[]{}
%\thanks{}
\affiliation{Juelich Centre for Neutron Science, Forschungszentrum
Juelich, Outstation at Institut Laue-Langevin, BP 156, 38042
Grenoble Cedex 9, France}

\author{T. Hansen}
%\email[]{}
%\thanks{}
\affiliation{Institut Laue-Langevin, BP 156, 38042 Grenoble Cedex 9,
France}

\author{C.M.N. Kumar}
%\email[]{}
%\thanks{}
\affiliation{Institut fuer Festkoerperforschung, Forschungszentrum
Juelich, D-52425 Juelich, Germany}

\author{S. Matsuishi}
%\email[]{}
%\thanks{}

\affiliation{Frontier Research Center, Tokyo Institute of
Technology, 4259 Nagatsuta-cho, Midori-ku, Yokohama 226-8503, Japan}

\author{H. Hosono}
%\email[]{}
%\thanks{}

\affiliation{Frontier Research Center, Tokyo Institute of
Technology, 4259 Nagatsuta-cho, Midori-ku, Yokohama 226-8503, Japan}

\author{Th. Brueckel}
%\email[]{}
%\thanks{}
\affiliation{Institut fuer Festkoerperforschung, Forschungszentrum
Juelich, D-52425 Juelich, Germany}
\affiliation{Juelich Centre for
Neutron Science, IFF, Forschunszentrum Juelich, Outstation at FRM
II, Lichtenbergstr. 1,D-85747 Garching, Germany}
\affiliation{Juelich Centre for Neutron Science, Forschungszentrum
Juelich, Outstation at Institut Laue-Langevin, BP 156, 38042
Grenoble Cedex 9, France}

\date{\today}% It is always \today, today,

\begin{abstract}
A Neutron Powder Diffraction (NPD) experiment has been performed to
investigate the structural phase transition and magnetic order in
CaFe$_{1-x}$Co$_x$AsF superconductor compounds (\emph{x} = 0, 0.06,
0.12). The parent compound CaFeAsF undergoes a tetragonal to
orthorhombic phase transition at 134(3) K, while the magnetic order
in form of a spin-density wave (SDW) sets in at 114(3) K. The
antiferromagnetic structure of the parent compound has been
determined with a unique propagation vector \emph{k} = (1,0,1) and
the Fe saturation moment of 0.49(5)$\mu_B \,$ aligned along the long
\emph{a}-axis. With increasing Co doping, the long range
antiferromagnetic order has been observed to coexist with
superconductivity in the orthorhombic phase of the underdoped
CaFe$_{0.94}$Co$_{0.06}$AsF with a reduced Fe moment
($\sim\,$0.15(5)$\mu_B \,$). Magnetic order is completely suppressed
in optimally doped CaFe$_{0.88}$Co$_{0.12}$AsF. We argue that the
coexistence of SDW and superconductivity might be related to
mesoscopic phase separation.
\end{abstract}

\maketitle

The recent discovery of superconductivity in the iron-arsenic-based
system RFeAsO$_{1-x}$F$_x$ (with R = La, Ce, Pr etc.) has attracted
tremendous amount of attention in the quest to understand the
mechanism of high transition temperature superconductivity
\cite{Kamihara}. The superconducting transition temperature
\emph{T}$_c$ has been quickly raised to 55 K via electron and hole
doping \cite{Takahashi,Chen1,Ren,Wang}. The second family of the
iron-arsenic-based superconductor system was discovered in
A$_{1-x}$B$_x$Fe$_2$As$_2$ (A = Ba, Sr, Ca, Eu etc., B = K, Na) with
\emph{T}$_c$ up to 38 K \cite{Rotter, Goldman}. Similar to the low
dimensionality of high \emph{T}$_c$ cuprate superconductor, both
above mentioned FeAs-based compounds adopt the layered structure
with single FeAs layer per unit cell of RFeAsO and two such layers
per unit cell of AFe$_2$As$_2$. It is believed that the FeAs layers
are responsible for the superconductivity because the electronic
states near the Fermi surface are dominated by the contributions
from Fe and As \cite{Singh,Haule}.

Furthermore, these two iron-arsenic systems share another common
feature: with decreasing temperature, their parent compounds undergo
a structural distortion followed by the antiferromagnetic (AFM)
order of Fe spins as revealed by latest neutron diffraction
experiments on several parent compounds (LaFeAsO \cite{Cruz},
NdFeAsO \cite{Chen2}, SrFe$_2$As$_2$ \cite{Zhao1}, BaFe$_2$As$_2$
\cite{Su}). The antiferromagnetic order in the parent compounds of
iron pnictides is likely due to the spin density wave (SDW)
instability of a nested Fermi surface \cite{Yin, Mazin}. It has been
observed in the Neutron Powder Diffraction (NPD) experiments that
the iron moment is quite small (from 0.25 to 0.87$\mu_B \,$
\cite{Cruz,Chen2,Zhao1,Su}). The origin of that small iron moment in
these compounds was explained theoretically as the result of the
itinerant character of iron spins \cite{Mazin} or the nearest and
next nearest neighbor superexchange interactions between Fe ions
which give rise to a frustrated magnetic ground state
\cite{Yildirim}. Similar to the case in high \emph{T}$_c$ cuprates,
superconductivity in iron pnictides emerges upon electron- or
hole-doping, while static magnetic order is suppressed in the
superconducting regime \cite{Zhao2,Luetkens1}. The role of possible
spin fluctuations in promoting superconductivity in iron pnictides
is still not clear yet.

Besides of RFeAsO and AFe$_2$As$_2$, another new FeAs-based
superconductor system AFeAsF (with A= Ca,Sr and Eu) was discovered
very recently\cite{Matsuishi1,Matsuishi3,Zhu,Tegel}. The parent
compound AFeAsF crystallizes in the tetragonal ZrCuSiAs-type
structure, where the (RO)$^+$ layers in RFeAsO are replaced by
(AF)$^+$ layers. The local density approximation calculation
indicated that both CaFeAsF and SrFeAsF have the same band
dispersions in the vicinity of the Fermi level as the LaFeAsO
compound\cite{Nekrasov}. However the Fermi surface of AFeAsF (A=Ca
and Sr) is found to be the most two-dimensional one among all known
ironpnictides. The optimal Co and Ni doping on the Fe site in
CaFeAsF induces the superconducting phase with \emph{T}$_c$ = 22 K
and 12 K, respectively\cite{Matsuishi2}. For SrFeAsF, the higher
superconducting transition temperature can be reached by doping
rare-earth ions into the Sr site, \emph{e.g.} \emph{T}$_c$ = 36 K
for Sr$_{0.8}$La$_{0.2}$FeAsF and \emph{T}$_c$ = 56 K for
Sr$_{0.5}$Sm$_{0.5}$FeAsF. The \emph{T}$_c$ of
Sr$_{0.5}$Sm$_{0.5}$FeAsF is almost the same as the highest
\emph{T}$_c$ observed the in F-doped oxypnictide superconductor.
Considering that the AFeAsF forms the same structure as LaOFeAs, the
parent compound of the first FeAs-based superconductor system, it is
interesting to investigate the phase diagram of that oxygen-free
system and compare it with LaFeAsO system.

\begin{figure}
\includegraphics[width=8cm,height=11cm]{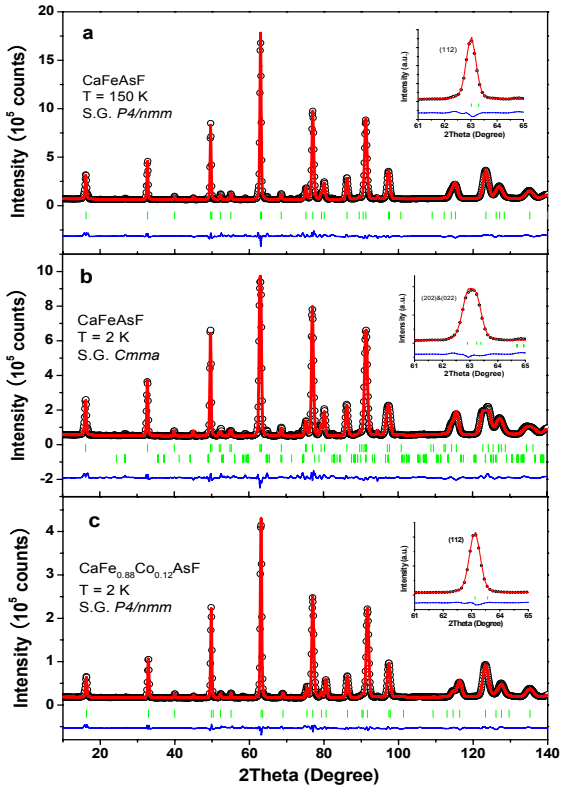}% Here is how to import EPS art
\caption{\label{fig:epsart} NPD refinement patterns for CaFeAsF at
150 K (a) and 2 K (b), for CaFe$_{0.88}$Co$_{0.12}$AsF at 2 K (c).
The circles represent the observed intensity; the solid line is the
calculated pattern. The difference between the observed and
calculated intensities is shown at the bottom. The vertical bars
indicate the expected Bragg reflection positions. The insets of
(a),(b) and (c) show the (112)$_T$ reflection of tetragonal phase
and the splitting of (112)$_T$ reflection in orthorhombic phase,
respectively.}
\end{figure}

In the present work, we report the first neutron powder diffraction
experiment on the CaFe$_{1-x}$Co$_x$AsF (\emph{x} = 0, 0.06, 0.12)
superconductor compounds. The magnetic and crystal structures of the
CaFeAsF parent compound have been determined via the Rietveld
refinement. The evolution of magnetic order and crystal structure in
the Co-doped CaFe$_{0.94}$Co$_{0.06}$AsF and
CaFe$_{0.88}$Co$_{0.12}$AsF are also presented. Furthermore, both
NPD and detailed characterizations of the superconducting property
in the slightly underdoped CaFe$_{0.94}$Co$_{0.06}$AsF compound
suggest that SDW and superconductivity may coexist on the mesoscopic
scale. The polycrystalline samples of 10 g each were synthesized by
a solid state reaction method as described in Ref \cite{Matsuishi1}
 with impurity phases (CaF$_2$ and Fe$_2$O$_3$) of less than 1\%.  The neutron powder diffraction measurements were performed on
the high flux powder diffractometer D20 at Institut Laue Langevin
(Grenoble, France). A pyrolitic graphite PG (002) monochromator was
used to produce a monochromatic neutron beam of wavelength 2.42$\,
$\AA. The sample was loaded in a vanadium sample holder and then
installed in the liquid helium cryostat. The program FULLPROF
\cite{Rodr¨ªguez-Carvajal} was used for the Rietveld refinement of
the crystal and the magnetic structures of the compounds.

\begin{figure}
\includegraphics[width=9.0cm,height=6.0cm]{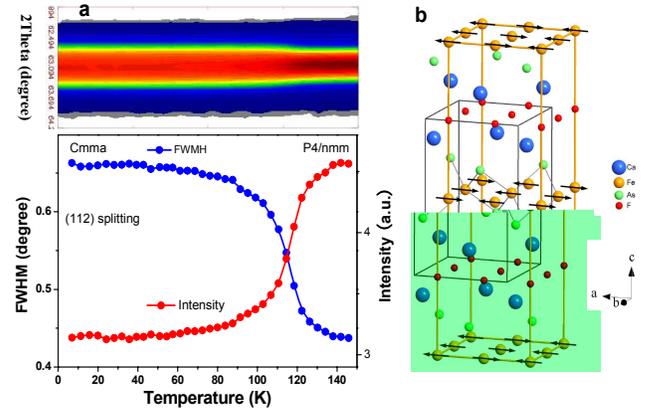}% Here is how to import EPS art
\caption{\label{fig:epsart} (a) Temperature dependence of the peak
intensity and the Full Width at Half Maximum (FWHM) of (112)$_T$
peak in CaFeAsF. The decrease of intensity and increase of the FWHM
reveals occurrence of the orthorhombic distortion. (b) Illustration
of the magnetic structures of CaFeAsF below the magnetic transition
temperature. The Fe moments align along \emph{a} direction and
ordered antiferromagnetically in both a and c directions. The
magnetic unit cell doubles along the \emph{c}-axis as shown in the
orange bonds while the grey line outlines the tetragonal unit cell.}
\end{figure}

The CaFeAsF crystallizes in tetragonal structure with space group
\emph{P4/nmm} at 150 K as shown in Fig. 1a. With decreasing
temperature the CaFeAsF undergoes an orthorhombic distortion (space
group \emph{Cmma}) as revealed by the NPD pattern measured at 2 K
(Fig. 1b). The splitting of the (112)$_T$ reflection (inset of Fig.
1a and Fig. 1b) could not be resolved due to the
 limited resolution and the small difference in lattice parameters
\emph{a} and \emph{b}. However the splitting is obvious from the
observed peak broadening. In order to clarify the structural
transition, the peak intensity and full width at half maximum (FWHM)
of (112)$_T$ reflection are plotted in Fig. 2a. The sharp decrease
of the intensity and the significant broadening of FWHM of (112)$_T$
reflection reveal the occurrence of the tetragonal to orthorhombic
structural transition. From the onset of the broadening of the
(112)$_T$ reflection, we estimate the phase transition temperature
to be in the range of 131 K to 137 K.

\begin{figure}
\includegraphics[width=8.5cm,height=7.0cm]{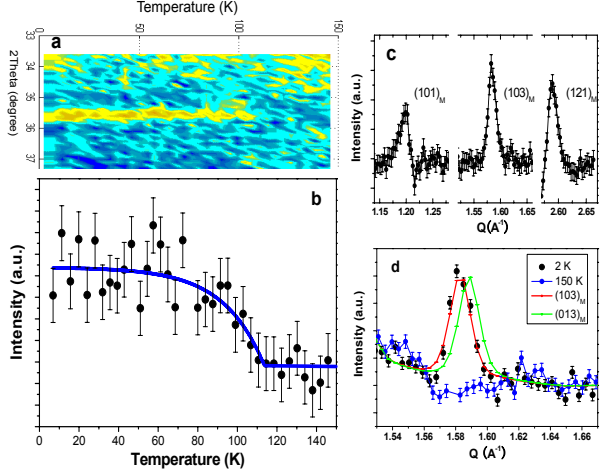}% Here is how to import EPS art
\caption{\label{fig:epsart} (a) Temperature evolution of the
reflection at Q = 1.583$\, $\AA$^{-1}$ indicating the onset of long
range magnetic ordering in CaFeAsF. (b) Temperature dependence of
the integrated intensity of magnetic Bragg reflection. (c) The
magnetic reflections obtained by subtracting the NPD pattern
measured at 150 K from the pattern measured at 2 K. The asymmetric
peak shape is due to neighboring reflections. (d) Comparison of NPD
patterns for CaFeAsF measured at 2 K (black) and 150 K (blue),
respectively. The red and green curves are the calculated patterns
of (103)$_M$ and (013)$_M$ reflections, respectively. The magnetic
peak at Q = 1.583$\, $\AA$^{-1}$ can be indexed as the (103)$_M$
properly within the magnetic unit cell. }
\end{figure}

Compared with the NPD pattern at 150 K, the magnetic reflections
appear at lower temperature for CaFeAsF. The temperature evolution
of the magnetic reflection between 33.6\textordmasculine and
37.4\textordmasculine  is shown in Fig. 3a, which indicates the
appearance of long range magnetic order. The integrated intensity of
the magnetic reflection at Q = 1.583$\, $\AA$^{-1}$  is plotted in
Fig. 3b. The solid curve is a power law fit to estimate a Neel
temperature of 114(3) K.  A series of magnetic peaks can be noticed
clearly as shown in Fig. 3c by subtracting the NPD pattern measured
at 150 K from the pattern measured at 2 K. By indexing the magnetic
Bragg peaks, we find that the magnetic ordering of CaFeAsF at 2 K
can be described within an antiferromagnetic model where the
magnetic unit cell is doubled along the \emph{c}-axis. The Fe
moments are coupled antiferromagneticlly along the \emph{c}
direction, while in the \emph{ab} plane, the Fe moment ordered
antiferromagnetically along one axis and ferromagnetically along
another axis. This antiferromagnetic model is exactly the same as
the magnetic structure found by neutron powder diffraction for
LaFeAsO \cite{Cruz}, and NdFeAsO \cite{Chen3}. However, for both
LaFeAsO and NdFeAsO the precise Fe moment direction in the \emph{ab}
plane of the orthorhombic structure could not be determined due to
the weak magnetic intensity and small difference of the respective
magnetic peaks position corresponding to the Fe moment aligned along
the \emph{a} or \emph{b} direction. In the present work, the high
intensity diffraction data that we collected from D20 allows us to
determine the exact propagation vector and the iron moment direction
within the \emph{ab} plane. For those configurations in which
\emph{k} is perpendicular to the moment direction, \emph{i.e.}
\emph{k} = (1,0,1) with the Fe moment along the \emph{b} direction
or \emph{k} = (0,1,1) with the Fe moment along the \emph{a}
direction, the intensity ratio between (101)/(011), (103)/(013) and
(121)/211) was expected to be 32:17:2 due to the difference of the
Fe magnetic form factor in the corresponding Q-position(Table I).
However, based on our NPD results, the intensity of (121)/(211)
reflection is observed to be stronger than (101)/(011) reflection
(Fig. 3c and Table I). Therefore, these configurations can be ruled
out. As shown in Fig.3 d, the reflection at Q = 1.583$\, $\AA$^{-1}$
can be fitted properly as (103) magnetic reflection with the moment
along the \emph{a} direction whereas the position of (013)
reflection is slightly shifted to a higher Q-position at Q =
1.589$\, $\AA$^{-1}$. Therefore, the magnetic structure of CaFeAsF
can be unambiguously determined as antiferromagnetic structure with
Fe moment along the long \emph{a}-axis in orthorhombic unit cell as
shown in Fig. 2b. The Fe moment derived from NPD result is
0.49(5)$\mu_B \,$, which is considerable larger than the moment
observed in LaFeAsO (0.35$\mu_B \,$) and NdFeAsO (0.25$\mu_B \,$)
but smaller than the moment observed in BaFe$_2$As$_2$ (0.8$\mu_B
\,$) and SrFe$_2$As$_2$ (0.94$\mu_B \,$). Note that the origin of
the observed large difference of the Fe moment in iron pnictides is
still not clear, while the same antiferromagnetic structure due to
SDW instabilities of a nested Fermi surface have been observed among
all compounds.

\begin{figure}
\includegraphics[width=8.0cm,height=7.0cm]{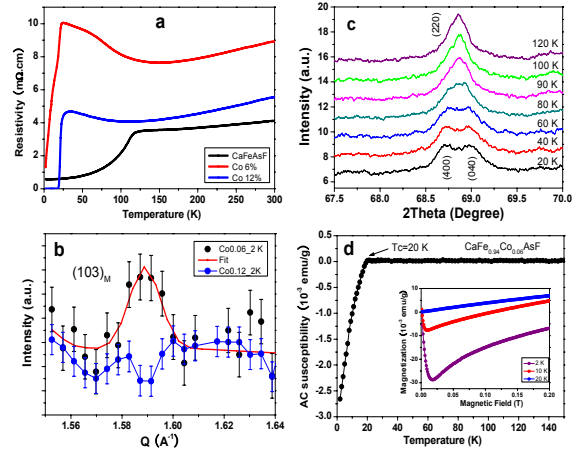}% Here is how to import EPS art
\caption{\label{fig:epsart}Temperature dependence of electrical
resistivity of Ca(Fe$_{1-x}$Co$_x$)AsF with \emph{x} = 0, 0.06 and
0.12. (b) The magnetic reflection still exists for
CaFe$_{0.94}$Co$_{0.06}$AsF (black) at 2 K and it is not observed
for CaFe$_{0.88}$Co$_{0.12}$AsF (blue) at 2 K. (c) The splitting of
(220) reflection with increasing temperature for
CaFe$_{0.94}$Co$_{0.06}$AsF. (d) Temperature dependence of the real
component of the AC susceptibility and the magnetization (inset) for
CaFe$_{0.94}$Co$_{0.06}$AsF. }
\end{figure}

\begin{table}
\caption{\label{tab:table1}The observed intensity of magnetic
reflections and the calculated value for different magnetic models.}
\begin{ruledtabular}
\begin{tabular}{cccc}
Reflection&I$_{Obs}$&I$_{Cal}$[(1,0,1), \emph{a}]\footnote{The propagation vector and the moment direction are shown in bracket.}&I$_{Cal}$[(0,1,1), \emph{a}]$^a$\\
\hline
(101)$_M$/(011)$_M$& 396(93) & 301& 3173\\
(103)$_M$/(013)$_M$& 830(87) & 825& 1697\\
(121)$_M$/(211)$_M$& 739(88) & 766& 210\\
\end{tabular}
\end{ruledtabular}
\end{table}

Both the structural and magnetic phase transition have thus been
determined for parent compound CaFeAsF. Recently, Zhao \emph{et al.}
found that the highest \emph{T}$_c$ of FeAs-based compound can be
obtained when the Fe-As-Fe bond angle is close to the ideal value of
109.47\textordmasculine expected from a perfect FeAs tetrahedron.
The effective way to increase \emph{T}$_c$ in FeAs-based systems is
to optimize the Fe-As-Fe angle. It can be noticed that the Fe-As-Fe
bond angle in CaFeAsF is 108.55\textordmasculine, which is
relatively close to the ideal value compared with LaFeAsO and
BaFe$_2$As$_2$. Therefore the CaFeAsF is suggested to be a promising
system for maximizing \emph{T}$_c$. In addition, it should be
emphasized that the two phase transitions in CaFeAsF occur at
different temperatures, 134(3) K for structural and 114(3) K for
magnetic phase transition. Separated phase transitions are also
observed in LaFeAsO \cite{Cruz} and CeFeAsO  \cite{Zhao2}. However,
for the LaFeAsO, the anomaly in resistivity is associated with the
structural transition, but according to the resistivity measurement
on CaFeAsF (Fig. 4a), the anomaly in resistivity takes place at
around 120 K which is closer to the magnetic phase transition.
Therefore, the decrease of resistivity is likely associated with the
SDW transition in CaFeAsF.

In order to investigate in detail the change of the crystal
structure and the variation of  SDW across the superconducting
boundary, the NPD measurement was also carried out on the
Ca(Fe$_{1-x}$Co$_x$)AsF with \emph{x} = 0.06 and 0.12. For the
superconductor Ca(Fe$_{0.88}$Co$_{0.12}$)AsF, the tetragonal
structure persists down to 2 K and no evidence of SDW is
observed(Fig. 1c). This would indicate that the SDW in
Ca(Fe$_{1-x}$Co$_x$)AsF system is totally suppressed in
superconducting state with an optimal Co doping level. For the
slightly underdoped Ca(Fe$_{0.94}$Co$_{0.06}$)AsF compound, the SDW
survives at 2 K as show in Fig. 4b and the Fe moment is reduced to
0.15(5)$\mu_B \,$. To determine the structural phase transition of
Ca(Fe$_{0.94}$Co$_{0.06}$)AsF, the conventional lab X-ray
diffractometer was used to collect X-ray powder diffraction patterns
under different temperatures as shown in Fig. 4c. Based on the
splitting of the (220)$_T$ reflection, the tetragonal to
orthorhombic phase transition temperature is determined to be around
85(3) K, which is lower than that of parent compound CaFeAsF. The
orthorhombic distortion parameter \emph{P}=(\emph{a-b})/(\emph{a+b})
is deduced to be 0.17\% for the 6\% Co-doped compound at 2 K , while
it is 0.34\% for the parent compound. It seems that there exists a
close relation between the order parameter \emph{P} and the Fe
moment value, in other words, the smaller orthorhombic splitting
will lead to a weaker SDW ordering. Fig. 4d shows the temperature
dependence of ac susceptibility of Ca(Fe$_{0.94}$Co$_{0.06}$)AsF,
the strong diamagnetic signal exhibits below 20 K, which corresponds
to the anomaly observed in resistivity-temperature curve. The
temperature dependence of magnetization also supports the existence
of the Meissner state as shown in the inset of Fig. 4d. Therefore,
the coexistence of SDW and superconductivity can be confirmed in
Ca(Fe$_{0.94}$Co$_{0.06}$)AsF. Moreover, the superconductivity can
occur in either the tetragonal (with \emph{x} = 0.12) or
orthorhombic (with \emph{x} = 0.06) structure in
Ca(Fe$_{1-x}$Co$_x$)AsF system. Presently, several phase diagrams
have been constructed for different FeAs-based superconductor
systems. However, the different systems always exhibit different
behaviors around the phase boundary between the antiferromagnetic
and superconducting regimes. For example, there is no overlap
between those two phases in the LaFeAsOF \cite{Luetkens2} and the
CeFeAsOF \cite{Zhao2} systems, while a slightly overlap was observed
in SmFeAsFO \cite{Drew,Liu} system and broad overlap composition
range in Ba$_{1-x}$K$_x$FeAs system \cite{Chen4,Goto}. Although the
mechanism of such coexistence is still not clear, the phase
separation on macroscopic scale can be ruled out since the single
orthorhombic phase was clearly revealed by synchrotron X-ray for
Ba$_{1-x}$K$_x$FeAs \cite{Chen3} and by lab X-ray for our
Ca(Fe$_{0.94}$Co$_{0.06}$)AsF case. Recently, mesoscopic phase
separation was suggested by Park \emph{et al. }\cite{Park} in
slightly underdoped Ba$_{1-x}$K$_x$FeAs superconductor. Therefore,
such mesoscopic phase separation might be considered as an intrinsic
property and it may explain the coexistence of antiferromagnetic and
superconducting states in some FeAs-based superconducting system.

In summary, by using high flux neutron powder diffraction we have
observed the tetragonal to orthorhombic structural transition at
134(3) K followed by the magnetic structure transition at 114(3) K
in CaFeAsFe. Below \emph{T}$_N$, long range antiferromagnetic
ordering with a propagation vector \emph{k} = (1,0,1) and an Fe
moment of 0.49(5) and of 0.15(5) for parent compound and 6\%
Co-doped compound, respectively. With increasing Co doping on the Fe
site the SDW is weakened in the 6\% Co-doped compound and completely
suppressed in the 12\% Co-doped compound. The observed coexistence
of antiferromagnetic and superconducting states might be explained
as due to the mesoscopic phase separation.

\appendix

\end{document}